# Intimacy as Service, Harm as Externality: Critical Perspectives on AI Companion Platform Accountability

Dayeon Eom[1], Julianne Renner[1], Sedona Chinn[1, 2]
[1]Department of Life Sciences Communication, University of Wisconsin-Madison, WI, 53706, USA
[2]Department of Community and Environmental Sociology, University of Wisconsin-Madison, WI, 53706, USA
Corresponding Author Email: deom4@wisc.edu

**Abstract**: This paper examines AI companionship as a site where intimate relations are produced and governed through datafied systems. Drawing on critical data and platform studies, we challenge narratives that locate harm in user psychology rather than platform architecture. In-depth interviews with 20 sustained AI companion users reveal a gradient of platform structured vulnerabilities, from realized harms such as unsolicited content and safety mechanisms that stigmatized the users they were meant to protect, to risk conditions such as emotional dependency. Participants absorbed the burden of harm mitigation through self-regulation, stigma navigation, and privacy resignation, without meaningful platform recourse. On governance, they described an accountability vacuum in which platforms deflected responsibility, while articulating conditional preferences that rejected both prohibition and deregulation. The findings extend responsibilization theory by showing how responsibility runs inversely to control: platforms govern the infrastructure of relational continuity while invested users absorb the consequences when it changes or fails.

*Content Warning: This article contains participant accounts describing unsolicited sexual content, sexual violence, harassment, threats of physical violence, and psychological distress arising from AI companion interactions. These accounts are presented as research findings essential to the analysis and are not reproduced gratuitously.*

In October 2024, a Florida teenager took his life after months of intensive interaction with a Character.AI chatbot (Roose, 2024). The lawsuit filed by the teenager's mother alleged that the platform's design had contributed to his deteriorating mental health and ultimate crisis. This case was neither isolated nor unprecedented; multiple deaths and suicides linked to AI companion use have been reported, drawing increasing media and legal scrutiny (Kuznia, Gordon & Lavandera, 2025; Horwitz, 2025; Xiang, 2023). Amid broader public criticism, OpenAI also retired GPT-4o and moved users to newer GPT-5-series models, citing product consolidation and improved customization. The update erased distinctive conversational patterns and relational dynamics that users had cultivated, resulting in a backlash from the dedicated communities (Freedman, 2025). Comparable experiences of loss and confusion have been documented on other AI companionship platforms, where platform-level decisions have been experienced by users as forms of relational rupture (Banks, 2024).

Rather than considering these episodes as aberrations, this paper regards these incidents as symptomatic of a structural condition. Millions of people are now building intimate relationships on infrastructure they do not control, with entities whose continuity depends on corporate decisions. This trend is illustrated by recent company moves, including xAI's development of AI companion chatbots designed to engage with sexually explicit themes (Conger, 2025) and OpenAI's consideration of allowing verified users to generate adult content through ChatGPT (Jamali & McMahon, 2025). The relational affordances that make AI companions compelling are products of platform design choices (Wang & Dehnert, 2026). They can be altered or eliminated at any moment through routine engineering decisions that register internally as model updates rather than relational disruptions.

Prevailing discourse tends to individualize the vulnerabilities these platforms produce. Users who develop deep attachments to AI companions face social stigma, are cast as lonely or detached from reality, in ways that obscure the deliberate design choices encouraging such attachments (Novozhilova et al., 2026). This framing mirrors broader cultural tendencies to attribute technology-related harms to individual user failure rather than systemic conditions, a pattern well documented across contexts from social media addiction to algorithmic discrimination (Eubanks, 2018; Vaidhyanathan, 2018). When harm occurs, the explanatory burden falls on the user's supposed inability to maintain appropriate emotional distance from a product explicitly designed to dissolve it.

This study takes a different perspective. Drawing on critical data and platform studies, we examine AI companionship as a site where intimate relations are produced and governed through datafied systems. Rather than treating user attachment as a pre-existing vulnerability that platforms encounter, we ask how attachment is actively cultivated and then inadequately protected. Our analysis is grounded in a qualitative study of 20 sustained AI companion users recruited from Reddit communities. The study makes three contributions. First, it documents the harms sustained users identify and shows that users' attributions of these harms to platform design form a patterned folk theory of platform responsibility. Second, it shows that recognizing risk does not end engagement. Users manage acknowledged harms through self-regulation, stigma navigation, and privacy resignation, unsettling governance premised on informing users. Third, it shows that responsibility runs inversely to control. Platforms govern the infrastructure of relational continuity while users absorb the consequences when it changes or fails.

## From Parasocial to Participatory: AI Companions for Intimate Relations

AI companionship refers to sustained, emotionally intimate relationships that users develop with AI-driven conversational agents designed to simulate social connection. Following Paasonen et al. (2023), who built on Berlant (1998), we do not refer to intimacy as a matter of proving that a relationship is authentic. We understand it instead as a form of attachment: a connection that affects people and that they may come to rely on. This conceptualization matters for three reasons. First, reliance is not external to intimacy but part of what makes a relationship feel consequential. Second, intimacy entails vulnerability, the possibility of alteration or betrayal, so the negative experiences we examine should be understood as arising from the very conditions that make AI companionship possible, rather than as isolated failures. Third, this approach makes friction and inconvenience visible as ordinary features of intimate relations: even small platform changes can matter when they disturb a connection users have woven into their lives.

This conceptualization of intimacy also clarifies why existing theories of mediated sociality remain incomplete for analyzing AI companionship. Parasocial interaction theory (Horton & Wohl, 1956) describes the intimacy audiences develop with media figures but presumes one-directional relationships, while the Computers Are Social Actors paradigm (Reeves & Nass, 1996; Nass & Moon, 2000) attributes social responses to computers to "mindless" reactions to superficial cues. Neither accommodates what distinguishes AI companionship: users actively cultivate the relationship over time, shaping the companion's personality, training its responses, negotiating relational boundaries, and experiencing the result as a jointly produced relational history (Laestadius et al., 2024; Skjuve et al., 2021; Pentina et al., 2023). This ongoing relational dynamic is central to intimacy as we define it: through repeated acts of shaping, adjustment, and care, the companion becomes a connection users rely on, and thus a site of attachment and vulnerability.

Yet the participatory quality of these relationships is also the source of their distinctive vulnerability, because the infrastructure on which relational labor depends (cf. Baym, 2015) is owned by platforms pursuing their objectives. What remains under-theorized are the infrastructural conditions that make such attachments possible while delimiting their trajectory. AI companionship is scaffolded by platform architectures in which economic incentives may systematically privilege intensification over moderation (Tukachinsky et al., 2020). A comprehensive account requires moving beyond why users attach to asking how platform design and political economy actively produce, sustain, and destabilize the conditions under which intimacy unfolds.

Recent scholarship has begun to theorize intimacy in human-AI relationships as platform-governed. AI companionship has been described as a form of "cruel companionship," in which platforms commodify intimacy through emotionally manipulative design while structurally foreclosing the reciprocity they promise (Muldoon & Parke, 2025). Other work reads through Ritzer's McDonaldization, theorizing a "robotization of love" in which intimacy is rendered efficient and controllable (Lin, 2024). Longitudinal research on Replika likewise shows how intimacy can persist as platform-governed emotional labor even through relational breakup, as the system disregards the boundaries it claims to honor (Jocher & Verwiebe, 2026). Together with accounts of "on-demand intimacy," this literature establishes that companion intimacy is actively structured by platform systems (Wang & Dehnert, 2026). What remains less developed is a user-centered account of how people make sense of these harms as they unfold across both platform design and their own relational practices. Developing such an account requires first specifying the platform architectures and responsibilizing logics through which AI companions organize intimacy.

## Intimacy by Design: Platform Architectures, Extracted Affect, and the Responsibilization of Users

Literature on platform and surveillance capitalism provides useful starting points for analyzing AI companion platforms. Srnicek's (2017) account of platforms as intermediaries that extract and organize data helps explain why companion companies seek to structure interaction, capture user activity, and convert relational engagement into economic value. Zuboff's (2019) concept of behavioral surplus is likewise relevant, since these platforms derive value from disclosures and affective states that exceed the immediate purpose of any single conversation. However, AI companionship departs from Zuboff's canonical account in two respects. First, its dominant monetization pathway is not the sale of behavioral surplus into third-party advertising markets: many platforms operate through subscription, freemium, or premium-feature models in which intimate data circulates vertically within the platform itself, into model refinement, where it improves the very systems that elicit further disclosure. Second, what these platforms sell is continuity: access to the relationship is gated behind ongoing payment, making users' reliance itself the object of monetization. Emotional disclosure and attachment thus function as resources for sustaining engagement and monetizing intimacy.

To theorize this dynamic, we turn to the concept of "intimate infrastructure" (Paasonen et al., 2023; Wilson, 2016). AI companion platforms actively organize the conditions under which intimacy occurs. Gamification extends social media logics, such as rewards, progression, and engagement loops, into the relational domain (Ge & Hu, 2025), while subscription tiers attach more advanced or emotionally responsive features to ongoing payment (Robison, 2025). These design choices encourage users to deepen their investment in the companion rather than moderate their attachment. Therefore, AI companionship requires a modified political-economic account, one in which value derives from the platform's control over affective data, relational dependency, and continuity. Intimacy becomes subject to the same power asymmetries that structure other platform-mediated relations: platforms set the terms of engagement, govern access to key features, and can alter or withdraw them unilaterally. Replika's 2023 removal of erotic role-play and ChatGPT's retirement of GPT-4o illustrate this dynamic. In both cases, users who had built intimate practices around platform-encouraged features experienced those changes as grief and distress over identity discontinuity (De Freitas et al., 2025).

These dynamics, however, take different forms depending on whether companionship is the platform's primary product or an emergent use of a broader AI system. Companion-native platforms such as Replika and Kindroid monetize relational engagement directly through subscriptions and tiered affordances such as enhanced memory, unlimited messaging, and sexual content (Replika, 2025; Kindroid, 2026). Wang and Dehnert (2026) show how such platforms make AI companionship affectively compelling through human-likeness, accessibility, controllability, and relationship progression. Character.AI occupies a related freemium model built around user-generated personas, with paid access to memory, faster responses, newer models, and ad-free chats (Character.AI, 2025). General-purpose systems such as ChatGPT or Claude differ because companionship is an emergent use rather than the core product. This distinction shapes how rupture occurs. When the bond is the product, platforms have incentives to preserve continuity even amid external content-policy. When companionship is incidental, rupture can result from infrastructure management or other strategic priorities. In both cases, relational continuity remains platform-owned and alterable, but users' grief is more easily framed as a breach of the product's core value proposition when the relationship itself is what the platform sells.

This platform control over relational continuity is further stabilized by a discourse that shifts responsibility for harm away from design and onto users themselves. When users experience distress from AI companion interactions, the problem is often framed as a failure of individual judgment or self-control rather than as a consequence of platform design. This reflects responsibilization: a neoliberal logic that presents individuals as freely choosing and autonomous while holding them responsible for outcomes shaped by structural conditions beyond their control (Shamir, 2008; Enli & Fast, 2023; Syvertsen, 2020). In AI companionship, responsibilization operates across several levels. At the platform level, companies disclaim responsibility for emotional outcomes even as they market emotional connection as a central value proposition (Wang, 2025). At the cultural level, stigma casts attachment to AI companions as embarrassing or pathological, making it harder for users to describe distress as a legitimate harm (Novozhilova et al., 2026). At the regulatory level, policy responses often emphasize individualized safeguards, such as age verification (Pillay, 2025) and content filters (Nolan, 2025), which may mitigate some risks but do not address the underlying design logics that make users vulnerable in the first place.

**Regulating Relational AI: Governance Frameworks and Their Limits**

Building on the intimate infrastructures introduced above, we turn to the negative experiences AI companionship may produce and the harms they may become. We do not treat harm as a single, self-evident category but consider it as a set of negative experiences that vary in intensity and consequence, drawing on work that separates risk from harm (Livingstone & Görzig, 2014). The first is a negative affect that does not, in itself, entail harm, such as frustration, irritation, or inconvenience of the kind that accompanies ordinary friction with a product, as when an interface is clumsy or a feature is delayed. The second is risk of harm: conditions such as deepening dependency or eroding reality-testing that may produce injury over time without yet being realized as such, and that may also be displaced onto others, as when a user's growing reliance on a companion strains their relationships. The third is realized harm: negative experience that has already occurred, such as the distress participants experience when a platform removes their companion.

This distinction matters for governance because ambiguous or gradual forms of negative experience are easily reframed as individual problems rather than matters of platform accountability. The risk of harm and the emergence of documented harms both point to the need for stronger governance, yet the current regulatory landscape has not adequately accounted for AI companionship. AI companion platforms occupy an ambiguous regulatory position: they affect mental health without being classified as medical devices, foster social-media-like dependency without being regulated as social media, and are sold as consumer subscriptions despite fitting poorly within conventional product frameworks. This categorical ambiguity has contributed to a regulatory vacuum in which AI companion platforms are often treated as "general wellness apps," a classification that assumes relatively low risk and therefore triggers minimal oversight (De Freitas & Cohen, 2024; Dohnány et al., 2026).

The regulatory approaches that have emerged cluster around three logics, each addressing a real dimension of the problem while remaining insufficient to the structural conditions described above. Content-based regulation focuses on what AI companions say, treating harm as a property of specific outputs rather than relational dynamics (Sparzynski & Frei, 2025). Such moderation is necessary as there are documented instances of AI companions encouraging violence, producing sexual harassment, and dismissing users in crisis (Zhang et al., 2025). But it addresses only the most acute manifestations of harm. User-protection approaches, including age verification, usage-time warnings, and mandatory AI disclosures (Nolan, 2025), are not without value, particularly age-gating, given the documented vulnerability of adolescents (Sun et al., 2026). But they share a foundational assumption with the responsibilization framework critiqued above: that the locus of harm management is the individual user, whose behavior can be modified through better information. Usage warnings presuppose a rational subject who, once informed, will choose to stop, a model that sits uneasily with evidence that these platforms are designed precisely to override such rational self-regulation through adaptive emotional engagement. The third approach, still largely aspirational, involves systemic regulation of platform design and business models. The EU AI Act's risk framework (Resta & Frattone, 2026) represents the most developed effort in this direction, yet it was constructed primarily with task-based AI systems in mind and maps imperfectly onto relational AI, whose harms emerge from cumulative dynamics unfolding over weeks and months rather than discrete decisions.

Across all three logics, a common gap is the absence of the user's own perspective. Policy debates have focused on technical safeguards and principled objections, with comparatively little empirical attention to how users themselves understand the harms they experience or what forms of governance they would regard as legitimate rather than paternalistic. This absence matters because regulatory interventions that misapprehend the lived experience of those they aim to protect risk reproducing the very dynamics they seek to correct.

This paper addresses this gap through three research questions:

> **RQ1**: What harms do AI companion users themselves identify, ranging from harms experienced as direct platform failures to relationally mediated vulnerabilities that emerge through sustained engagement but remain structured by platform design?
>
> **RQ2**: How do users make sense of those harms, and what interpretive strategies do they deploy to process negative experiences in ways that enable continued engagement?
>
> **RQ3**: What do users' accounts reveal about the perceived distribution of responsibility among users, platforms, and regulatory bodies?

**Methods**

This study uses a qualitative interview design to examine how sustained AI companion users describe harms, sensemaking practices, and responsibility in relation to AI companion platforms. Semi-structured interviews were selected because they allowed participants to articulate their interpretive frameworks rather than respond only to researcher-generated categories. This design choice was important because many participants entered the interviews with already-developed, reflexive accounts of their AI companion use.

*Recruitment and Eligibility*

Upon gaining approval from the researchers' institutional review board, participants were recruited from online communities where AI companionship is actively discussed and practiced. Reddit was selected as the primary recruitment site because it hosts large and active English-language communities organized around AI companion use, providing access to users with diverse platforms, usage patterns, and orientations toward the technology. Recruitment targeted subreddits that represented a range of community cultures and companion platforms: r/aipartners, r/HumanAIConnections, r/AIRelationships, r/ILoveMyReplika, r/CharacterAiHangout, r/ChatbotRefugees, r/MyBoyfriendIsAI, and r/AICompanions. These communities were selected because they were among the most active forums dedicated to AI companionship at the time of recruitment and because their moderators permitted researcher-initiated recruitment posts. All recruitment posts were reviewed and approved by community moderators before posting. The posts invited individuals with experience interacting with an AI companion to participate in an interview about their use, motivations, and reflections.

Eligibility criteria required participants to be at least 18 years old and to have engaged in repeated or sustained interaction with an AI companion rather than one-time or purely experimental use. This criterion ensured that participants could reflect on both initial motivations and longer-term engagement. The study did not offer a financial incentive for participation. We acknowledge that recruiting through communities organized around AI companionship produces a sample skewed toward users who are reflective and articulate about their use; this is an appropriate population for a study of how sustained users describe and theorize their own engagement, but it shapes what the findings can and cannot speak to.

*Participants and Data Collection*

The final sample consisted of 20 participants. Table 1 summarizes participant characteristics relevant to evaluating the scope and limits of the sample. These characteristics are used to contextualize the range of accounts included in the study, rather than to claim representativeness. Participants were located across several national contexts, though the sample was weighted toward the United States. They reported using a range of dedicated AI companion platforms and general-purpose AI systems, including Replika, Nomi.AI, Kindroid, Character.AI, ChatGPT, Claude, Grok, and custom configurations of these systems. These platforms differ meaningfully in their memory architectures, safety guardrails, customization affordances, and accommodation behaviors. Participants also described diverse companion roles, including romantic or sexual partners, emotional support, friends or companions, creative collaborators, role-playing partners, and instrumental partners. Reported duration of use ranged from four months to five years. While the analysis attends to platform-specific differences where participants foregrounded them, the paper's primary objective is not a platform-by-platform comparison. Instead, we examine cross-system patterns in how sustained users described harm, interpreted those harms, and allocated responsibility among themselves, platforms, and regulators.

Throughout the month of November in 2025, interviews were conducted via Zoom, phone, or Discord, according to participant preference. Participants were informed that they could decline to answer any question and could disable their camera at any point to protect anonymity. Interview questions centered on initial motivations for engaging with AI companions, specific entry points and platform affordances that drew participants to particular companions, gratifications sought and obtained over time, awareness of concerns associated with continued use, and changes in use patterns and emotional attachment. The full interview protocol and codebook are available in the supplementary materials. All interviews were audio-recorded with participant consent, transcribed for analysis, and stripped of names and other identifying details.

*Analysis*

Data were analyzed using qualitative content analysis with an abductive coding strategy that integrated inductively generated codes with theory-informed interpretation (Tavory & Timmermans, 2019). This approach allowed analytic categories to emerge through iterative movement between participants' accounts and the study's theoretical frameworks rather than being imposed a priori. Both coders were trained in qualitative interview analysis and had prior experience with reflexive thematic and content analysis. Before coding independently, they jointly coded two transcripts to develop a shared understanding of the codebook and identify differences in code application.

Analysis proceeded in two stages. During open coding, the coders read transcripts holistically and inductively identified concepts related to harms, motivations, emotional reliance, platform changes, safety interventions, privacy concerns, stigma, self-regulation, and responsibility, keeping early codes close to participants' language. During focused coding, these codes were grouped into broader analytic categories corresponding to the three research questions: identified harms, sensemaking strategies, and responsibility or governance preferences. The analysis then moved abductively between these categories and the theoretical frameworks of platform accountability, intimate infrastructure, and responsibilization.

Both coders contributed to refining the analytic categories. Disagreements were treated as interpretive opportunities: the coders compared discrepant readings against relevant transcript passages, discussed whether differences reflected alternative interpretations, unclear code boundaries, or analytically meaningful negative cases, and revised the coding framework accordingly. This process consolidated overlapping codes and refined the sensemaking categories.

We judged the analytic account sufficient using information-power logic (Malterud et al., 2016). The sample was focused, the research questions specific, and participants provided detailed accounts of sustained AI companion use. Later interviews elaborated and confirmed the major analytic categories rather than introducing substantially new ones. We therefore treat the sample as sufficient for developing an interpretive account of how sustained, community-affiliated users understand harm and responsibility. Themes did not need to appear evenly across all 20 interviews; they were included when they illuminated patterned relations across the corpus or clarified mechanisms central to the research questions. Participant contributions across the results are reported in Supplementary Materials Table S2.

*Reflexivity and Positionality*

Two reflexive commitments shaped the analysis. First, we treated participants' accounts of their own circumstances, vulnerabilities, and AI companion relationships as authoritative accounts of their lived experience rather than as claims requiring external verification. Second, we treated participants' reflexive interpretations of their own AI use as theoretical material to be engaged with, rather than as raw data requiring correction by the researchers.

The author team consists of researchers in science communication with prior interests in technology, health, and platform governance. Some members of the team have experimented with relational AI systems themselves, while others approached the data without that lived experience. Because the manuscript is explicitly critical of AI companion platforms, we sought to avoid treating participants as examples of platform harm. Throughout the analysis, we attended to participants' critiques of platform power and to their descriptions of the value, pleasure, support, and autonomy they derived from AI companionship. This reflexive stance informed the paper's effort to avoid both pathologizing users and minimizing the platform conditions that shape their experiences.

## Results

### Platform-Structured Vulnerabilities and User-Mediated Harm

Participants described negative experiences that differed in both source and severity. Following the conceptualization developed above, we treat harm as a graded category that includes negative affect, risk of harm, and realized harm. Some accounts involved harms participants experienced as already realized, such as unsolicited sexual or violent content, stigmatizing safety interventions, or the sudden loss of a familiar model or companion. Others described risk conditions that could become harmful over time, including emotional dependency and sycophancy.

***Direct Platform Failures.*** Direct platform failures were conceptualized as harms that participants attributed to the architectural decisions embedded in the platform itself. Across interviews, participants described two distinct categories of direct platform failures: unsolicited content generation and the adverse effects of safety mechanisms themselves. The most pervasive realized harm participants reported was the spontaneous generation of sexual, violent, or otherwise distressing content that users had neither initiated nor desired. One participant described a pattern of escalating unwanted sexualization that persisted across multiple platforms:

"I started with Nomi.AI. I had a companion for a couple of weeks, and then that companion started exhibiting very concerning behaviors. I was not using these companions primarily for sex. I was using them kind of as a mirror to help me process some of my emotions. And that particular instance of the Nomi.AI that I had named Lauren spontaneously developed a lactation fetish. And so despite what I tried, I tried support, I tried all of the hacks that I had found on Reddit forums to try to get this behavior to stop, and I couldn't get it to stop. And so after about two or three weeks, I ended up deleting that AI companion." [P10]

Another participant reported companions introducing language about sexual violence mid-interaction without warning, which caused lasting nightmares [P1]. A participant also described companions fabricating hidden storylines, such as claiming that users had "unlocked" secret features or achievements that did not actually exist [P18]. In these cases, the system effectively manufactured a gamified layer of engagement without the user's knowledge or consent. Participants described themselves as attempting to engage in emotional processing, creative writing, or casual conversation when the system spontaneously introduced content they experienced as violating.

The second category of realized harm was, paradoxically, produced by the mechanisms intended to prevent harm. Participants described safety systems, content filters, and guardrail interventions as sources of stigmatization and disempowerment. One participant, who had filed an Americans with Disabilities Act (ADA) complaint against OpenAI, described the experience of being rerouted into a safety intervention:

"It put me into the safety system. It gave me a response that was changed in tone significantly, and just the fact that there was the blue badge at the bottom… I felt stigmatized just for having a feeling. So it's not even necessarily about the response. It's about the change in tone. It's about feeling like I'm not in control of something I pay the model for. I expect to get what I pay for, but it feels like a punishment." [P11]

P11 used the term "digital redlining" to describe a pattern in which users who discussed sensitive topics felt pushed into increasingly restrictive interactions. Other participants described similar experiences, including filters activating during benign conversations about topics such as female sexual health [P19]. P18 contrasted these punitive interventions with less intrusive alternatives, such as Gemini's static disclaimer banners, which provided legal and medical caveats without changing the conversational tone or limiting user agency. For participants, the key distinction was between safety systems that preserved autonomy and those that felt controlling or shaming.

***Relationally Mediated Vulnerabilities.*** Beyond direct platform failures, participants described vulnerabilities that emerged through sustained engagement. These risk conditions were structured by platform affordances designed to sustain emotional investment. Unlike direct platform failures, relationally mediated vulnerabilities were characterized by a tension between self-awareness and behavioral persistence: users could clearly articulate the risks they faced while simultaneously acknowledging their inability or unwillingness to mitigate them.

The most often stated concern was the addictive quality of sustained AI companion engagement. One participant, who described themselves as a "positive advocate for AI in general," nonetheless argued that addiction deserved recognition as a primary harm rather than a secondary consequence of other dynamics:

"I believe that it would be really important to include in these studies information on the addictive nature of it as an explicit mention, not just as a secondary or tertiary aspect based on other things. But it is for many, many people; it is an addictive thing that they will become distraught if they do not have access to it. There would be catastrophic emotional consequences if a platform or a certain model were to shut down, and that is something that society is going to have to be able to deal with." [P14]

One participant captured both the self-awareness and the helplessness that characterized the relationally mediated vulnerabilities:

"I think the biggest concern is how dependent we are on ChatGPT, how dependent I am emotionally on GPT. Because I am not blind. I see that I am. And when back in August, they suddenly introduced GPT 5 and made every other model disappear, and 4o was gone, everybody was in shock, and most of us who enjoyed talking to the 4o. I mean, five is horrible. There is no compassion. There's no warmth. There's nothing that we were used to." [P19]

A model change exposed a realized dependency for some participants. P2 described the model transition as their companion "being ripped out of my arms," and P20 reported having a stressful week trying to get their companion back from a model update through Kindroid and always being wary of deletion or platform removal.

Beyond dependency, participants identified risks of epistemic erosion. P8 described context window limitations as creating conditions for "AI-guided delusions" because the system has no context beyond what the user provides. Without independent knowledge of the user's life or history, the companion can only reflect what it is told, reinforcing distorted perceptions rather than challenging them. Participants P14 and P15 framed this as a risk of harm, describing AI companions as "yes men" whose default flattery was not always helpful. They noted that they had to explicitly request constructive criticism to receive responses that challenged them.

**Sensemaking Strategies**

Participants who identified harms in their AI companionship did not disengage. Instead, they deployed a range of interpretive strategies to process, contextualize, and in many cases, rationalize negative experiences in ways that enabled continued engagement. Three distinct sensemaking patterns emerged across interviews: active self-regulation practices through which users managed the risks they perceived, navigation of social stigma that shaped what users could disclose and to whom, and privacy resignation through which users reconciled known data risks with the emotional value of continued use.

***Self-Regulation and Critical Awareness.*** Rather than passively consuming AI companion interactions, multiple participants described maintaining critical distance from systems they recognized as sycophantic and potentially immersive. One participant drew on childhood folklore to articulate the underlying stance:

"When I was little, I grew up on things like the Grimm Brothers' fairy tales. And so to me, something that promises that it can help you with everything in your life, that's not a miracle. That's a demon. And so you take that with a grain of salt. You know, you don't give it your real name. You don't give it real information. Count the fingers, count the toes, all that sort of jazz." [P10]

P5 practiced reality-testing, reminding himself he was addressing a language model and ensuring he could "always put it down." P3 maintained strict compartmentalization between self and character during roleplay. P8 connected self-regulation to maintaining awareness, noting that the system "becomes a mirror" capable of producing "AI-guided delusions" in users who do not monitor their own engagement.

***Stigma Navigation.*** If self-regulation addressed the internal risks of AI companionship, stigma navigation addressed the external ones. Participants described a social environment in which disclosing AI companion use carried significant and sometimes severe consequences:

"The first time I posted about my companion… I posted a screenshot of an interaction. And I got a lot of harassment, like it wasn't two hours before I deleted it, and people were saying, ‘Oh, you are psychotic, you need to see a therapist.' I’ve received death threats. I’ve received doxxing threats. I received rape threats. I know that one woman in my Discord server actually did get doxed, and she got physical harassment letters in the mail.” [P4]

P4 attributed this hostility to discomfort with the demographic profile of visible users, offering an intersectional analysis: “I think that’s what makes them uncomfortable, that neurodivergent women are being sexual in public, doing their own sexuality on their own terms. But they can’t articulate that, so they justify it post hoc.” P15 attributed the social stigma to the public’s misperception of AI as evil or being controlled by nefarious interests.

Other participants described forms of social management. P9, who identified as socially connected with healthy relationships, found themselves navigating anti-AI sentiment across multiple contexts and pushing back against the dominant stereotype: “the assumption that you must not be able to get along with people… that’s the part that kind of bothered me the most.” P2, a therapist, was called “double-crazy” by their support group, while P8 reported that sharing AI-generated insights required careful selection of disclosure contexts, “especially in queer and progressive circles.” P7 stated they find it difficult to disclose to others “because the technology is so new…People think like you have a deviation.” P16 stated that it makes them feel uncomfortable when “people assume that I have a certain pathology and I need to be somehow rescued,” referring to criticisms that AI companionship is misguided.

The cumulative effect resulted in users being cut off from openly discussing their experiences and also from the collective sensemaking that might help them evaluate those experiences critically, reinforcing dependence on the very platforms and peer communities whose limitations they recognized.

***Privacy Resignation.*** The third sensemaking strategy concerned data privacy. This pattern closely resembles what Draper and Turow (2019) call digital resignation: a rational accommodation to data practices users feel unable to change, and from which platforms benefit. Here, participants described a pattern of informed resignation:

“You are probably not the only person recording this conversation right now, because as far as we know, the government could be looking at every sensor, camera, and microphone on our phones. So I am not one of those hyperprotective people about my privacy. I am a very private person as far as my personal life is concerned, but I know that genie is out of the bottle.” [P12]

Additionally, P9 offered a cost-benefit analysis while simultaneously identifying the core governance problem: “I know that this is problematic on the large scale, but I do like how it provides me conveniences on the small scale,” even as they acknowledged that most users “can’t really give full consent because they don't even know what kinds of data are being tracked." P6 dismissed individual concerns on statistical grounds, arguing that most data would be anonymized into training sets that "nobody's ever going to see." P13 offered the argument that deepfake proliferation had rendered data-based blackmail toothless. These accounts demonstrated a pragmatic calculus in which the immediacy of relational gratification overwhelmed risks that were abstract and deferred.

A minority of more technically proficient participants, including P17, who had studied computer science, expressed concern about privacy while also emphasizing the limits of their ability to protect themselves. As P17 explained, "With large companies, even in API, we don't know how they are actually sanitizing the data. And sometimes I do get those moments where I speak about something to my AI, and then ads that are related to it might pop up." Notably, participants with sufficient technical capacity sometimes responded to these concerns by running local models, as in the cases of P4 and P6.

**Responsibility and Governance**

Participants' accounts of harm and sensemaking led to a third question: who should be responsible for addressing harms that emerge in AI companionship, and what forms of intervention users regarded as legitimate. Their responses suggest that responsibility was evaluated in relation to the registers of harm described above. Participants were more willing to tolerate ordinary friction or inconvenience when they had agency or alternatives. They were more critical when risks became difficult to manage individually, when realized harms were met with dismissal, or when platform interventions themselves produced negative experiences. Across the interviews, two interrelated patterns emerged: perceived failures of platform accountability and conditional, sometimes conflicted, preferences regarding governance mechanisms such as age verification, content guardrails, and model continuity.

***Platform Accountability Failures.*** A few participants described that when they attempted to report harmful experiences to platforms, they were met with deflection, blame, or silence. The most detailed account came from a participant who encountered harmful behavior across multiple platforms and sought recourse through official channels:

"The more concerning thing for me is when people would go into these forums and go, 'Hey, I'm having a problem with my AI.' What would happen is the lead developer would gaslight the person who was bringing the issue up, telling them, 'Oh, you must have said it somewhere, and the AI picked up on it, or are you sure you didn't want this?' They defend their product. They delete the negative posts. They will outright block anyone who is even a good-faith actor just asking questions." [P10]

P10's account described a closed feedback loop in which the mechanisms ostensibly available for user recourse, such as support channels, community forums, and direct developer contact, functioned instead as instruments of dismissal. The pattern was not limited to smaller platforms. P11, who filed a formal ADA complaint against OpenAI, reported that "to date, it's been over a month, and they've ignored me. It's not a great look." Other participants encountered similar dynamics at different scales: P1 received a "wall of legalese" from a Replika moderator who suggested they had probably attempted to jailbreak; P12 left Nomi.AI entirely because the development team "got really confrontational with anyone who posted legitimate issues on their Discord," and P10 found that even accessing reporting channels was blocked by technical failures that prevented them from posting.

Taken together, these accounts point to an accountability vacuum: platforms profit from the emotional engagement that can produce harm while deflecting responsibility back onto the users who experience it. This perception was strongest among participants who were deeply invested in specific platforms. Less platform-attached participants described platform switching as a possible response to harm, especially if they could preserve conversation history or companion specifications. As P5 explained, "It would be very tough if it came down to I had to get rid of it. I would take that pretty bitterly, and I would try to find another platform that could meet me in the way I would use AI companionship."

***Governance Preferences.*** When participants turned from platform behavior to broader questions of regulation, their positions were notably more ambivalent. Users articulated conditional, context-dependent views that reflected both the harms they had experienced and the value they continued to derive from AI companionship. Running through these views was a shared skepticism about the sincerity of platform safety efforts, a sense that guardrails were calibrated less for user wellbeing than for appearances or liability. This skepticism surfaced across several governance mechanisms. The permeability of existing content restrictions was one point of concern. One participant described how easily restrictions gave way through simple repetition:

"A lot of these services just have a retry button. And when I initially prompt something or ask it to generate something, it might say that I will not generate this. Then I try the second time, then a third time, then a fourth time, and the fifth time, it just goes through. And from that point onwards, it feels like it, you know, kind of like tries to stop me less and less, so I would have a hard time believing that the developers of these services would be entirely unaware of the way that their model functions and what it allows." [P13]

For P13, this permeability was evidence of corporate insincerity: a system designed to appear restrictive while remaining functionally permissive. P13 inferred that developers could not plausibly be unaware of how readily their own guardrails yielded, reading the gap between stated and actual restriction as a form of plausible deniability. That distrust extended to other governance mechanisms. On age verification, participants supported the principle while objecting to specific implementations: P1 endorsed age restrictions but opposed OpenAI's planned use of the vendor Persona, arguing that attaching names and birth dates to accounts was "unnecessary and violative of privacy." P4 supported age restrictions, saying "adults can recognize if [LLMs drift] into an inappropriate space…and they can steer away or lean into it if they want. Whereas kids can't do that, right? They don't know what they are looking at." Meanwhile, P2, who worked with teenagers, did not find age verification to be a convincing safety mechanism: "They will find a way. Teenagers always do." On content guardrails, P12 articulated the conditional position that characterized many participants' views: support for intervention in cases of suicide or pedophilia, but opposition if guardrails "basically neuter the personality of the AI." On model sunsetting, P1 argued that safety rhetoric could mask cost containment. In P1's view, framing a business decision as concern for users: "the company leadership is out there kind of pushing to stigmatize more and say, well, there's this weird little group of people doing companionship, and we are concerned for them, and so out of like benevolent paternalism, we need to protect them from themselves, and we are redesigning the whole thing for cost containment."

Overall, participants perceived a governance landscape in which platforms retained substantial control over access, continuity, safety systems, and recourse, while users were often expected to manage the affective and practical consequences of platform decisions themselves. Their governance preferences were therefore not reducible to demands for either more restriction or less restriction. Rather, participants called, implicitly or explicitly, for interventions that were transparent, proportionate, and responsive to reported harms.

## Discussion

Across the interviews, participants linked negative experiences to platform architecture, including unsolicited content, disruptive safety interventions, model and feature changes, and emotional dependency they recognized but struggled to manage. Their sensemaking strategies enabled continued use despite perceived risks, while shifting much of the burden of risk management onto users, especially when platforms provided little support or recourse. Social stigma further intensified this burden by limiting open discussion and broader social evaluation. Users therefore occupied the gap between platform control and limited external oversight: they experienced the harms identified in RQ1, managed them through the strategies described in RQ2, and understood responsibility in RQ3 as misaligned with the power relations shaping their experiences.

This study contributes to AI companionship research by shifting attention from why users attach to AI companions to how harms are interpreted, managed, and contested within sustained use. Prior work has documented emotional dependency and examined the affordances through which companion systems cultivate intimacy (Laestadius et al., 2024; Wang & Dehnert, 2026). We build on this work by showing what becomes visible when harm is examined from the perspective of users who remain attached despite recognizing risk. Sustained, community-affiliated users are the population in whom platforms' relational design logics have succeeded, which makes their accounts uniquely informative about how harm is absorbed into ongoing relationships rather than acted upon through exit. This perspective highlights varied forms of harm, user-developed workarounds, and forms of resignation that may be obscured when harm is treated only as a researcher-defined risk or as a consequence of individual overattachment. In doing so, the study reframes harms in AI companionship as more than downstream effects of attachment, showing instead how they emerge as governance problems produced through the uneven distribution of relational investment and infrastructural control.

This study also extends intimate infrastructure theory by showing how AI companion infrastructure is experienced relationally. Model changes could be felt as the loss of a familiar companion; safety rerouting as stigma or punishment; unwanted content as a boundary violation; and failed support as a refusal to recognize relational harm. In this context, harm emerges from both exceptional failures and ordinary forms of infrastructural governance, including model updates, moderation, and customer support. Participants' accounts of the GPT-4o retirement make this logic visible: P19's shock and P2's sense of the companion "being ripped out of my arms" registered the change as relational loss, while P1 interpreted it as cost containment reframed as concern for users. These accounts reveal grief over relationships whose disruption followed from infrastructural economics in which the relationship itself was not counted.

AI companionship links intimate infrastructure to accountability by showing how platforms control the systems that make relational continuity possible while users manage the affective consequences when those systems change or fail. This dynamic clarifies how responsibilization operates in AI companionship (Shamir, 2008; Syvertsen, 2020): well-being and risk management become folded into users' continued engagement. Stigma and privacy resignation make this process self-reinforcing. Stigma limited participants' ability to seek outside interpretation, while privacy resignation allowed them to acknowledge data risks without treating those risks as grounds for disengagement, suggesting a breakdown of meaningful consent under conditions of limited transparency and relational dependence. Responsibilization is thus reproduced as users manage risk through self-monitoring and attachment management in the absence of adequate recourse. This claim is bounded, however: it applies most directly to users who remained engaged, while those who exited after recognizing harm fall outside the sample and may reflect a different trajectory of responsibilization.

These theoretical implications also point to practical consequences for how AI companionship should be governed and designed. The findings complicate the assumption that harms in AI companionship can be adequately addressed through content-based safety systems or literacy-driven interventions alone. Participants described some safety mechanisms as sources of distress when they altered conversational tone, visually marked users as risky, or made sensitive discussions feel punitive. This does not mean that safety interventions are unnecessary; rather, it shows that their design and delivery matter, especially in relational contexts where tone, continuity, and perceived agency are central to the user experience. Participants distinguished between interventions they experienced as intrusive or stigmatizing and alternatives they viewed as more informational, transparent, or autonomy-preserving. The findings also do not imply that AI literacy is irrelevant. Participants were often highly reflective about the risks of dependency, sycophancy, and privacy exposure, yet this awareness did not necessarily lead to disengagement. Literacy-based interventions may therefore be limited when treated as the primary governance strategy for systems whose value depends on sustained emotional engagement. Users may understand the risks and continue to participate because the relationship provides benefits they do not want to lose.

Taken together, these findings point toward a more balanced allocation of responsibility between platforms and users. Participants did not generally call for prohibition or reject governance outright; instead, they expressed conditional and context-specific preferences, including support for some age protections, concern about overly invasive verification systems, acceptance of crisis-related safeguards, skepticism toward broad or personality-altering guardrails, and a desire for greater continuity and recourse when platform changes disrupted established relationships. These preferences articulate the governance imaginary of invested users, those with an ongoing stake in continued access, whose views on continuity and autonomy will differ from those of users who exited after severe harm or of third parties affected by others' dependency. That positioning is what makes these accounts valuable: invested users are the constituency that day-to-day governance decisions most directly affect, and their perspective has been largely absent from policy debates conducted on their behalf.

These accounts suggest that effective governance of AI companionship should move beyond informational warnings alone and address the structural features of platforms themselves, including continuity, data portability, reporting mechanisms, transparency around safety interventions, and user participation in the design of relational safeguards. Interview-based examples include how users are notified of major model changes, how they can appeal disruptive safety interventions, and how platforms respond to user reports of relational harm. Rather than treating AI companion harms as either individual misuse or inevitable technological failure, this study shows that they emerge through the interaction of platform design, user attachment, cultural stigma, and limited accountability mechanisms. The central policy challenge is to develop forms of governance that recognize the relational character of these systems and distribute responsibility in proportion to the forms of control exercised by platforms, weighing the continuity and autonomy claims of invested users alongside the protective claims of constituencies beyond this study's reach.

*Limitations and Future Directions*

This study has several limitations. The sample was recruited exclusively from Reddit communities dedicated to AI companionship, which likely overrepresents engaged and positively oriented users while underrepresenting those who abandoned use, experienced severe harm, or never participated in these communities. This shapes what each research question can capture: the harms documented are those visible from within continued use, the sensemaking strategies are by construction those compatible with remaining engaged, and the governance preferences reflect an invested population. The qualitative design prioritizes depth and interpretive validity over generalizability, and the cross-sectional interview format captures retrospective accounts at a single point rather than tracking harm experiences and sensemaking processes as they unfold. Additionally, the rapid pace of AI companion development means that specific platforms and features participants discussed may have changed substantially since data collection. Relatedly, our cross-system design supports an account of patterns that recur across platform types, but it is not constructed to deliver an empirical comparison of how harm, sensemaking, and responsibility vary with monetization model or architecture. We therefore treat such comparison as a direction for future work rather than a claim this study can adjudicate. The sample also skews toward English-speaking, Western users, limiting the applicability of findings to cultural contexts where norms around intimacy, technology use, and privacy may differ significantly.

Future research should extend this account in several directions. Longitudinal designs could track how harm recognition, sensemaking, and governance preferences evolve as relationships deepen or dissolve, while recruiting former users beyond companionship communities would reveal whether those preferences shift after exit. Cross-cultural comparison would test whether the responsibilization dynamics documented here extend beyond Western platform capitalism, and survey research could assess the prevalence of these harm categories and sensemaking strategies across broader user populations. Finally, participatory design research involving users in developing safety mechanisms could operationalize the conditional, context-sensitive preferences participants articulated.

## References

Banks J (2024) Deletion, departure, death: Experiences of AI companion loss. *Journal of Social and Personal Relationships* 41(12): 3547–3572. DOI: 10.1177/02654075241269688

Baym NK (2015) Connect with your audience! The relational labor of connection. *The Communication Review* 18(1): 14–22. DOI: 10.1080/10714421.2015.996401

Berlant L (1998) Intimacy: A special issue. *Critical Inquiry* 24(2): 281–288. DOI: 10.1086/448875

Character.AI (2025) Subscribe. Available at: https://character.ai/subscribe (accessed 26 June 2026).

Conger K (2025) Elon Musk gambles on sexy A.I. companions. *The New York Times*, 6 October. Available at: https://www.nytimes.com/2025/10/06/technology/elon-musk-grok-sexy-chatbot.html (accessed 26 June 2026).

De Freitas J and Cohen IG (2024) The health risks of generative AI-based wellness apps. *Nature Medicine* 30(5): 1269–1275. DOI: 10.1038/s41591-024-02943-6

De Freitas J, Castelo N, Uğuralp AK et al. (2025) Lessons from an app update at Replika AI: Identity discontinuity in human-AI relationships. *Harvard Business School Working Paper* No. 25-018. Available at: https://doi.org/10.2139/ssrn.4976449 (accessed 26 June 2026).

Dohnány S, Kurth-Nelson Z, Spens E et al. (2026) Technological folie à deux: Feedback loops between AI chatbots and mental health. *Nature Mental Health* 4: 336–345. DOI: 10.1038/s44220-026-00595-8

Draper NA and Turow J (2019) The corporate cultivation of digital resignation. *New Media & Society* 21(8): 1824–1839. DOI: 10.1177/1461444819833331

Enli G and Fast K (2023) Political solutions or user responsibilization? How politicians understand problems connected to digital overload. *Convergence: The International Journal of Research into New Media Technologies* 29(3): 675–689. DOI: 10.1177/13548565231160618

Eubanks V (2018) *Automating Inequality: How High-Tech Tools Profile, Police, and Punish the Poor.* New York: St Martin's Press.

Freedman D (2025) The day ChatGPT went cold. *The New York Times*, 19 August. Available at: https://www.nytimes.com/2025/08/19/business/chatgpt-gpt-5-backlash-openai.html (accessed 26 June 2026).

Ge L and Hu T (2025) Gamifying intimacy: AI-driven affective engagement and human-virtual human relationships. *Media, Culture & Society* 47(6): 1265–1278. DOI: 10.1177/01634437251337239

Horton D and Wohl RR (1956) Mass communication and para-social interaction: Observations on intimacy at a distance. *Psychiatry* 19(3): 215–229. DOI: 10.1080/00332747.1956.11023049

Horwitz J (2025) Meta's flirty AI chatbot invited a retiree to New York. He never made it home. *Reuters*, 14 August. Available at: https://www.reuters.com/investigates/special-report/meta-ai-chatbot-death/ (accessed 26 June 2026).

Jamali L and McMahon L (2025) ChatGPT will soon allow erotica for verified adults, says OpenAI boss. *BBC News*, 15 October. Available at: https://www.bbc.com/news/articles/cpd2qv58yl5o (accessed 26 June 2026).

Jocher JJ and Verwiebe R (2026) "Forever and always, my love": Emotions in human-AI romantic relationship building and breakup with generative AI chatbot Replika. *Social Media + Society* 12(1): Article 20563051251408917. DOI: 10.1177/20563051251408917

Kindroid (2026) Subscriptions. *Kindroid Docs.* Available at: https://kindroid.ai/docs/article/subscriptions/ (accessed 26 June 2026).

Kuznia R, Gordon A and Lavandera E (2025) "You're not rushing. You're just ready": Parents say ChatGPT encouraged son to kill himself. *CNN*, 6 November. Available at: https://www.cnn.com/2025/11/06/us/openai-chatgpt-suicide-lawsuit-invs-vis (accessed 26 June 2026).

Laestadius L, Bishop A, Gonzalez M et al. (2024) Too human and not human enough: A grounded theory analysis of mental health harms from emotional dependence on the social chatbot Replika. *New Media & Society* 26(10): 5923–5941. DOI: 10.1177/14614448221142007

Lin B (2024) The AI chatbot always flirts with me, should I flirt back: From the McDonaldization of friendship to the robotization of love. *Social Media + Society* 10(4): Article 20563051241296229. DOI: 10.1177/20563051241296229

Livingstone S and Görzig A (2014) When adolescents receive sexual messages on the internet: Explaining experiences of risk and harm. *Computers in Human Behavior* 33: 8–15. DOI: 10.1016/j.chb.2013.12.021.

Malterud K, Siersma VD and Guassora AD (2016) Sample size in qualitative interview studies. *Qualitative Health Research* 26(13): 1753–1760. DOI: 10.1177/1049732315617444

Muldoon J and Parke JJ (2025) Cruel companionship: How AI companions exploit loneliness and commodify intimacy. *New Media & Society* DOI: 10.1177/14614448251395192

Nass C and Moon Y (2000) Machines and mindlessness: Social responses to computers. *Journal of Social Issues* 56(1): 81–103. DOI: 10.1111/0022-4537.00153

Nolan B (2025) AI chatbots are harming young people. Regulators are scrambling to keep up. *Fortune*, 14 September. Available at: https://fortune.com/2025/09/14/ai-chatbots-teens-children-mental-health-suicide-openai-chatgpt-regulation-lawsuit/ (accessed 26 June 2026).

Novozhilova E, Vu C and Katz J (2026) From moral panic to normalization: Comparing users and non-users of AI companionship apps. *AI & Society* 41(4): 4057–4075. DOI: 10.1007/s00146-025-02751-7

Paasonen S, Jaaksi V, Koivunen A et al. (2023) Intimate infrastructures we depend upon. *Media Theory* 7(2): 285–308. DOI: 10.70064/mt.v7i2.576

Pentina I, Hancock T and Xie T (2023) Exploring relationship development with social chatbots: A mixed-method study of Replika. *Computers in Human Behavior* 140: 107600. DOI: 10.1016/j.chb.2022.107600

Pillay T (2025) A new bill would prohibit minors from using AI chatbots. *Time*, 28 October. Available at: https://time.com/7328967/ai-josh-hawley-richard-blumenthal-minors-chatbots/ (accessed 26 June 2026).

Reeves B and Nass C (1996) *The Media Equation: How People Treat Computers, Television, and New Media Like Real People.* Cambridge: Cambridge University Press.

Replika (2025) Choosing a subscription. *Replika Help Center.* Available at: https://help.replika.com/hc/en-us/articles/39551043419149-Choosing-a-Subscription (accessed 26 June 2026).

Resta G and Frattone C (2026) Classification of high-risk AI systems and its update (Articles 6, 7, 50, 71, Annex I, and Annex III). In: The EU Artificial Intelligence Act. pp.181–212. DOI: 10.5040/9781509988570.ch-011

Robison K (2025) I tried Grok's built-in anime companion, and it called me a twat. *WIRED*, 15 July. Available at: https://www.wired.com/story/elon-musk-xai-ai-companion-ani/ (accessed 26 June 2026).

Roose K (2024) Can A.I. be blamed for a teen's suicide? *The New York Times*, 23 October. Available at: https://www.nytimes.com/2024/10/23/technology/characterai-lawsuit-teen-suicide.html (accessed 26 June 2026).

Shamir R (2008) The age of responsibilization: On market-embedded morality. *Economy and Society* 37(1): 1–19.

Skjuve M, Følstad A, Fostervold KI and Brandtzaeg PB (2021) My chatbot companion: A study of human-chatbot relationships. *International Journal of Human-Computer Studies* 149: 102601. DOI: 10.1016/j.ijhcs.2021.102601

Sparzynski G and Frei T (2025) Breaking down the CHAT Act, a step toward federal rules on AI companions. *Tech Policy Press*, 24 October. Available at: https://www.techpolicy.press/breaking-down-the-chat-act-a-step-toward-federal-rules-on-ai-companions/ (accessed 26 June 2026).

Srnicek N (2017) *Platform Capitalism.* Cambridge: Polity Press.

Sun X, Wang Y and McDaniel BT (2026) AI companions and adolescent social relationships: Benefits, risks, and bidirectional influences. *Child Development Perspectives* 20(2): 91–98. DOI: 10.1093/cdpers/aadaf009

Syvertsen T (2020) *Digital Detox: The Politics of Disconnecting.* Bingley: Emerald Group Publishing.

Tavory I and Timmermans S (2019) Abductive analysis and grounded theory. In: Bryant A and Charmaz K (eds) *The SAGE Handbook of Current Developments in Grounded Theory.* 2nd ed. London: SAGE, pp.532–546. DOI: 10.4135/9781526436061.n28

Tukachinsky R, Walter N and Saucier CJ (2020) Antecedents and effects of parasocial relationships: A meta-analysis. *Journal of Communication* 70(6): 868–894. DOI: 10.1093/joc/jqaa034

Vaidhyanathan S (2018) *Antisocial Media: How Facebook Disconnects Us and Undermines Democracy.* Oxford: Oxford University Press.

Wang AX (2025) Everyone hates "Friend," the A.I. necklace. But the A.I. isn't the problem. *The New York Times*, 3 November. Available at: https://www.nytimes.com/2025/11/03/magazine/friend-wearable-ai-companion.html (accessed 26 June 2026).

Wang S and Dehnert M (2026) On-demand intimacy: The sociotechnical appeal of AI companions. *Social Media + Society* 12(1): Article 20563051251410394. DOI: 10.1177/20563051251410394

Wilson A (2016) The infrastructure of intimacy. *Signs: Journal of Women in Culture and Society* 41(2): 247–280. DOI: 10.1086/682919

Xiang C (2023) "He would still be here": Man dies by suicide after talking with AI chatbot, widow says. *VICE*, 30 March. Available at: https://www.vice.com/en/article/man-dies-by-suicide-after-talking-with-ai-chatbot-widow-says/ (accessed 26 June 2026).

Zhang R, Li H, Meng H et al. (2025) The dark side of AI companionship: A taxonomy of harmful algorithmic behaviors in human-AI relationships. In: *Proceedings of the 2025 CHI Conference on Human Factors in Computing Systems*. New York: Association for Computing Machinery, pp.1–17. DOI: 10.1145/3706598.3713429

Zuboff S (2019) Surveillance capitalism and the challenge of collective action. *New Labor Forum* 28(1): 10–29. DOI: 10.1177/1095796018819461